\documentclass[sigconf]{acmart}
\AtBeginDocument{%
  }

\setcopyright{acmlicensed}
\copyrightyear{2018}
\acmYear{2018}
\acmDOI{XXXXXXX.XXXXXXX}
\acmConference[Conference acronym 'XX]{Make sure to enter the correct
  conference title from your rights confirmation email}{June 03--05,
  2018}{Woodstock, NY}
\acmISBN{978-1-4503-XXXX-X/2018/06}

\usepackage{booktabs}

\usepackage{tcolorbox}
\definecolor{c1}{cmyk}{0,0.6175,0.8848,0.1490}
\definecolor{c2}{cmyk}{0.1127,0.6690,0,0.4431}
\definecolor{c3}{cmyk}{0.3081,0,0.7209,0.3255}
\definecolor{c4}{cmyk}{0.6765,0.2017,0,0.0667}
\definecolor{c5}{cmyk}{0,0.8765,0.7099,0.3647}

\definecolor{lightgrey}{rgb}{0.93,0.93,0.93}

\newtcbox{\hlprimarytab}{on line, rounded corners, box align=base, colback=c3!10,colframe=white,size=fbox,arc=3pt, before upper=\strut, top=-2pt, bottom=-4pt, left=-2pt, right=-2pt, boxrule=0pt}
\newtcbox{\hlsecondarytab}{on line, box align=base, colback=red!10,colframe=white,size=fbox,arc=3pt, before upper=\strut, top=-2pt, bottom=-4pt, left=-2pt, right=-2pt, boxrule=0pt}
\newcommand{\dashifted}{\raisebox{0.5\depth}{\tiny$\downarrow$}}
\newcommand{\uashifted}{\raisebox{0.5\depth}{\tiny$\uparrow$}}
\newcommand{\da}[1]{{\footnotesize\hlsecondarytab{\dashifted{#1}}}}
\newcommand{\ua}[1]{{\footnotesize\hlprimarytab{\uashifted{#1}}}}

\usepackage{multirow}
\usepackage{enumitem}
\usepackage{siunitx}
\usepackage{pifont}

\usepackage[framemethod=tikz]{mdframed}
\mdfdefinestyle{customstyle}{
  linecolor=gray!100,
  linewidth=3pt,
  innerleftmargin=3pt,
  topline=false,
  rightline=false,
  bottomline=false,
  leftline=true,
  innerrightmargin=3pt,
  innertopmargin=3pt,
  innerbottommargin=3pt,
  backgroundcolor=gray!15
}

\newenvironment{custommdframed}
  {\begin{mdframed}[style=customstyle]}
  {\end{mdframed}}

\begin{document}

\title{AdaptiveLLM: A Framework for Selecting Optimal Cost-Efficient LLM for Code-Generation Based on CoT Length}

\author{Junhang Cheng, Fang Liu$^\ast$, Chengru Wu, Li Zhang}
\thanks{$^{\ast}$Corresponding author.}
\affiliation{%
\institution{State Key Laboratory of Complex \& Critical Software Environment, School of Computer Science and Engineering \\ Beihang University, Beijing, China}
  \country{}
}

\email{chengjunhang7@gmail.com, {fangliu, 23230618, lily}@buaa.edu.cn}

\renewcommand{\shortauthors}{Cheng et al.}

\begin{abstract}
While Large Language Models (LLMs) have significantly advanced code generation efficiency, they face inherent challenges in balancing performance and inference costs across diverse programming tasks. Dynamically selecting the optimal LLM based on task difficulty and resource constraints offers a promising approach to achieve an optimal balance between efficiency and performance. However, existing model selection methods are resource-intensive and often neglect cost efficiency. Moreover, these approaches rely on human-annotated difficulty labels that are frequently inaccessible in real-world settings and may not align with the LLM's own assessment of task difficulty.
In this paper, we introduce AdaptiveLLM, a framework that dynamically selects optimal LLMs for a given coding task by automatically assessing task difficulty. Our framework first estimates task difficulty using Chain-of-Thought lengths generated by reasoning model, clusters these into three difficulty levels via k-means, and fine-tunes CodeBERT to embed difficulty-aware features. A trained XGBoost classifier then selects the best model for each problem, optimizing the performance-cost trade-off.
Experimental results show that AdaptiveLLM achieves a 7.86\% improvement in pass@1 score while reducing resource consumption by 88.9\% compared to baseline method ComplexityNet. When compared to a single model, AdaptiveLLM demonstrates an approximately 15\% accuracy improvement, while maintaining the same level of cost consumption. Apart from that, the difficulty assessment using CoT provides more reliable selection criteria than human evaluation.
Our replication package is available at \url{https://github.com/cjhCoder7/AdaptiveLLM}.

\end{abstract}




\maketitle

\begin{figure}[h]
	\centering
        \setlength{\abovecaptionskip}{0.1cm}
	\includegraphics[width=1\linewidth]{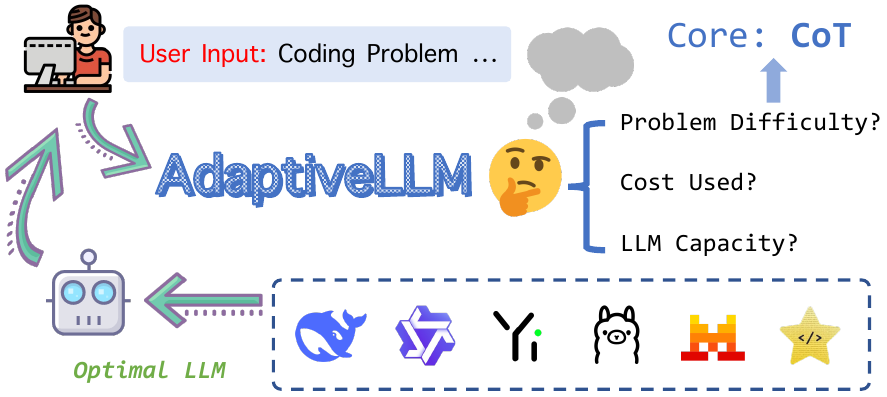}
	\caption{The Use of AdaptiveLLM.}
	\label{fig:the_use_of_adaptiveLLM}
        \vspace{-0.3cm}
\end{figure}

\section{Introduction}
Large Language Models have emerged as transformative tools in code understanding and generation, driving significant advancements in programming efficiency through intelligent assistants such as GitHub Copilot \cite{GitHub-Copilot} and Cursor \cite{Cursor}. These systems leverage LLM's contextual reasoning capabilities to predict and auto-complete code snippets, reducing development time and effort. However, expanding application scenarios reveal two critical challenges: \ding{182} LLM exhibit varying performance across programming tasks of different complexities, and \ding{183} their high computational costs for inference and training necessitate optimized resource allocation. 
Specifically, SWE-Bench \cite{jimenez2023swe} has demonstrated that different types of code LLMs possess varying capabilities in solving different types of code problems. Similarly, code LLMs with different parameter sizes also show distinct abilities in addressing code problems of varying difficulties. While larger models generally exhibit stronger problem-solving capabilities, smaller models can achieve comparable results on simpler tasks. Given that larger models incur higher operational costs and are overqualified for simpler problems, \textit{dynamically selecting the optimal LLM based on task complexity and resource constraints presents a promising strategy to balance efficiency and performance}.

To address this challenge, researchers have developed various model selection approaches, which generally fall into two categories. The first category comprises router-based approaches \cite{chen2024routerdc,jiang2023llm,chen2023frugalgpt,gupta2024language,yue2023large}. For example, RouterDC \cite{chen2024routerdc} introduces a novel routing mechanism that differs from traditional selection methods by embedding the problem-LLM matching into a vector space. This approach allows for a single LLM invocation to solve the problem, reducing resource consumption compared to previous strategies that required generating and screening responses from all LLMs in advance. However, RouterDC overlooks the difficulty levels of the problems themselves and relies solely on model responses during the framework construction to create problem-LLM matches, which can be resource-intensive. 
The second category selects different LLMs based on the difficulty levels of problems \cite{bae2023complexitynet,jeong2024adaptive,mallen2022not}. For instance, ComplexityNet \cite{bae2023complexitynet} uses difficulty tags to guide the selection of optimal models. However, this approach overlooks cost considerations during model invocation, frequently relying on high-performance but expensive closed-source models like GPT-3.5 and GPT-4.
Additionally, ComplexityNet relies on annotated difficulty tags, which are often unavailable in real-world settings, and they may not accurately reflect an LLM's intrinsic perception of task difficulty.
Furthermore, directly estimating difficulty using LLMs has proven unreliable due to the randomness in their predictions.

Recent advances in reasoning models have opened new possibilities for automating problem difficulty assessment \cite{guo2025deepseek-r1,team2025kimi,gpto1,gpto3}. These models produce step-by-step Chains-of-Thought (CoT) reasoning process, where longer reasoning sequences often correlate with higher problem complexity. This method not only mimics human cognitive patterns but also provides a systematic framework for evaluating difficulty. By integrating reasoning models into the framework and considering model costs, we can potentially overcome the limitations of both Router-based and difficulty-based methods for code generation tasks.

To this end, we propose AdaptiveLLM, a framework that dynamically selects optimal code generation models based on automated difficulty assessment. 
AdaptiveLLM first estimates the difficulty of each coding problem according to the CoT length generated from LLMs with enhanced reasoning capabilities. These CoT lengths are clustered into three difficulty levels using k-means, and the resulting labels are used to fine-tune the CodeBERT embedding model \cite{feng2020codebert}. This fine-tuning process enriches the problem embeddings with difficulty-aware features.
Subsequently, we train an XGBoost classifier to select the best-performing model for each problem, considering both response quality and resource efficiency to generate a balanced ranking. By dynamically matching problems with the most suitable models, AdaptiveLLM optimizes the trade-off between performance and cost. \textbf{Experimental results on three datasets of varying difficulty show that AdaptiveLLM achieves a 7.86\% improvement in capability and an 88.9\% reduction in cost compared to baseline methods.}

Our contributions are summarized as follows:
\begin{itemize}[left=0pt]
    \item We present AdaptiveLLM, a framework for selecting the most cost-effective LLM based on problem difficulty, model capability, and model cost. It enables personalized selection of the optimal LLM for each problem.
    \item We propose a novel programming task difficulty assessment method based on CoT of reasoning models. This automated approach eliminates the need for human intervention by using the length of CoT to estimate problem complexity.
    \item We evaluate AdaptiveLLM and the baseline method on benchmark datasets. Results show that AdaptiveLLM achieves superior performance with significantly lower resource consumption. Additionally, our analysis of the CoT difficulty assessment method reveals that CoT more accurately reflects LLMs' perception of problem difficulty compared to human labels.
\end{itemize}

\section{Related Work}
\subsection{LLMs for Code Generation}
Large Language Models have achieved significant breakthroughs in natural language processing, particularly demonstrating exceptional capabilities in code understanding and generation \cite{guo2025deepseek-r1,achiam2023gpt,liu2024deepseek,touvron2023llama,team2025kimi,team2024gemma,team2024gemini,grattafiori2024llama,yang2024qwen2}. Previous works focused on fine-tuning pretrained LLMs to address tasks of varying types and complexities. For example, mathematical reasoning tasks have been addressed through specialized fine-tuning of models such as WizardMath \cite{luo2023wizardmath}, Qwen2.5-Math \cite{yang2024qwen2.5-math} and MetaMath \cite{yu2023metamath}. Similarly, code generation capabilities have been improved through fine-tuning, as demonstrated by models such as Qwen2.5-Coder \cite{hui2024qwen2} and DeepSeek-Coder-V2 \cite{zhu2024deepseek}. These models exhibit significant performance gains in their respective target domains. 

Code generation benchmarks have evolved to assess LLM's capabilities across complexity levels. HumanEval \cite{chen2021evaluating} is a popular benchmark, whitch then extended to multilingual adaptations \cite{orlanski2023measuring,cassano2022multipl,zheng2023codegeex}, novel code completion paradigms \cite{muennighoff2023octopack}, and enhanced testing frameworks \cite{liu2023your}. High-complexity evaluation leverages CodeContests \cite{li2022competition} for algorithmic challenges and SWE-Bench \cite{jimenez2023swe} for real-world software engineering scenarios. FullStackBench \cite{liu2024fullstack} further enables cross-domain evaluation spanning 16 programming languages with SandboxFusion execution.

Our study selects three datasets: HumanEval, LeetCodeSample, and CodeContests. Additionally, we choose eight code LLMs and DeepSeek R1 reasoning model.

\subsection{Evaluation of Coding Problems}
The difficulty of programming problems is a crucial factor in selecting the optimal model. 
Various approaches have been proposed for assessing the difficulty of programming problems.

LeetCodeSample \cite{liu2024fullstack} and CodeContests \cite{li2022competition} extracted from competitive programming platforms, such as LeetCode \cite{leetcode} and Codeforces \cite{codeforce}, often rely on user performance data to quantify problem difficulty. However, these metrics suffer from cross-platform incomparability and dependence on user performance data.
Another approach involves analyzing the complexity of solutions. \citet{wang2024selection} evaluated problem difficulty using five code complexity metrics: Line Complexity, Cyclomatic Complexity \cite{ebert2016cyclomatic}, Halstead Complexity \cite{hariprasad2017software}, Cognitive Complexity \cite{campbell2018cognitive}, and Maintainability Index \cite{welker2001software}. However, not all the problems have standard solutions. Furthermore, for problems with multiple valid solutions, the variability in code complexity across different implementations will influence it.
\citet{bae2023complexitynet} and \citet{jeong2024adaptive} evaluate difficulty based on the number of interactions between the problem and a model to achieve a correct solution. It does not differentiate well between high level problems, because no matter how many iterations are done it fails to generate the correct code.

In this study, we propose a novel approach to evaluate problem difficulty based on the CoT length generated by reasoning LLMs.

\begin{figure*}[h]
	\centering
         \setlength{\abovecaptionskip}{0.1cm}
	\includegraphics[width=1\linewidth]{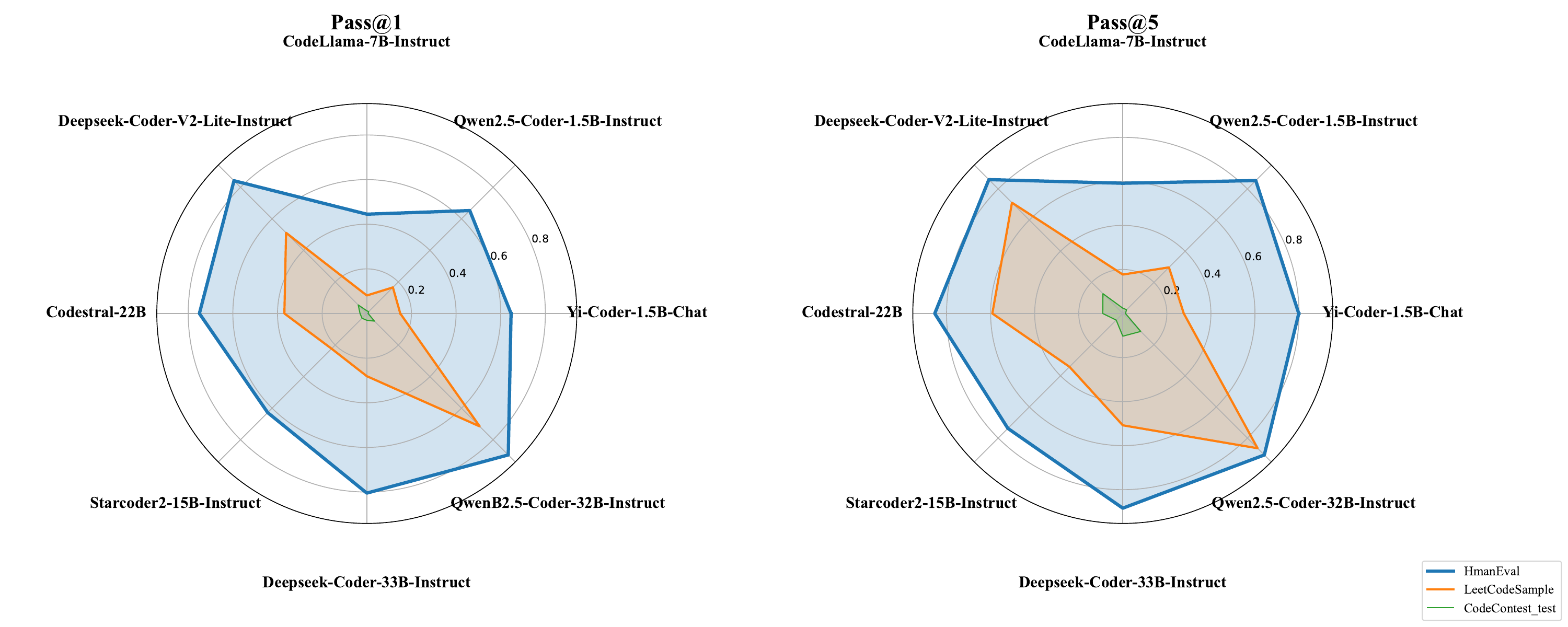}
	\caption{The performance of eight code LLMs on HumanEval, LeetCodeSample and CodeContests.}
	\label{fig:Dataset_Score}
\end{figure*}

\subsection{Model Selection}
Different LLMs exhibit distinct strengths and weaknesses and no single LLM currently dominates across all tasks. For such issues, the prevalent solution currently is to select the optimal LLM based on the specific task.

\citet{jiang2023llm} introduced PairRanker and GenFuser, which generate improved outputs by synthesizing results from all LLMs. However, this method requires invoking LLMs $O(T^2)$ times, where $T$ represents the number of models.
To optimize both performance and efficiency in LLM selection, researchers have proposed various methods. Cascading strategies \cite{chen2023frugalgpt,gupta2024language,yue2023large} sequentially invoke a series of pre-ranked models, typically ordered by capacity from smallest to largest. This process stops when a model's output meets a predefined confidence threshold. Nevertheless, these methods still requires at least $O(T)$ model invocations during inference.
Building on these efforts, \citet{chen2024routerdc} proposed RouterDC, which employs a routing mechanism to precisely identify the optimal model for a given task. Meanwhile, \citet{bae2023complexitynet} introduced ComplexityNet, a method that involves fine-tuning an embedded model, DaVinci-002 \cite{brown2020language}, to select models. It also requires only a single call to the selected model to complete the task.

However, model selection is not only influenced by LLM's capability but also by additional factors such as computational cost and problem difficulty.
So, we propose AdaptiveLLM, which integrates multiple factors, including model performance, computational cost, and problem difficulty, into its decision-making process.

\section{Preliminary Study}
\label{sec:preliminary-study}

\begin{figure}[h]
	\centering
        \setlength{\abovecaptionskip}{0.1cm}
	\includegraphics[width=1\linewidth]{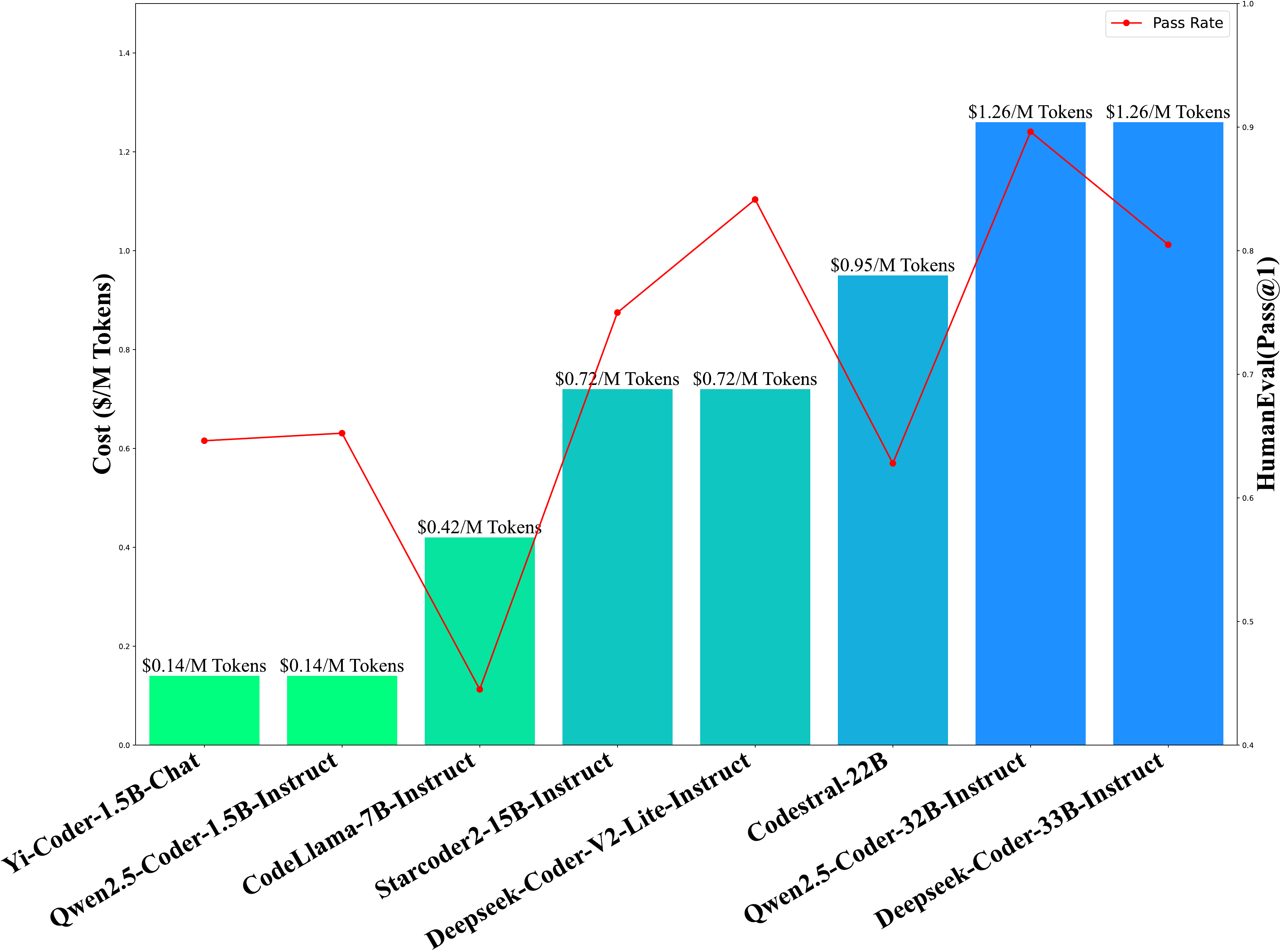}
	\caption{The relationship between Cost and Performance.}
	\label{fig:Cost_Pass@1_Compare}
        \vspace{-0.3cm}
\end{figure}

To investigate the capabilities differences among various LLMs and compare their capabilities against pricing, we conduct a preliminary study in this section.
We select eight code LLMs: Yi-Coder-1.5B-Chat \cite{young2024yi}, Qwen2.5-Coder-1.5B-Instruct \cite{hui2024qwen2}, CodeLlama-7B-Instruct \cite{roziere2023code}, Starcoder2-15B-Instruct \cite{lozhkov2024starcoder}, DeepSeek-Coder-V2-Lite-Instruct \cite{zhu2024deepseek}, Codestral-22B \cite{codestral}, DeepSeek-Coder-33B-Instruct \cite{guo2024deepseek} and Qwen2.5-Coder-32B-Instruct \cite{hui2024qwen2}. All these models are evaluated on three benchmark datasets: HumanEval \cite{chen2021evaluating}, LeetCodeSample \cite{liu2024fullstack}, and CodeContests \cite{li2022competition}. Figure \ref{fig:Dataset_Score} illustrates the evaluation results, while Figure \ref{fig:Cost_Pass@1_Compare} shows the comparison of cost (\$/M Tokens) and HumanEval pass@1 score across eight models. Our key findings are summarized below.

\textbf{Dataset Difficulty Analysis}: The evaluated datasets exhibit a clear gradient of complexity: \textbf{HumanEval < LeetCodeSample < CodeContests}. HumanEval focuses on fundamental programming tasks with 164 Python problems that requires function-level code completion based on signature and functional descriptions. LeetCodeSample, sourced from the LeetCode, presents more complex algorithmic challenges with longer problem descriptions. CodeContests, derived from Codeforces competitions, which problems involving advanced algorithmic paradigms such as dynamic programming, graph theory, and combinatorial optimization. And These problems have strict time/space constraints and require full-file code generation rather than function completion. Additionally, CodeContests includes challenges with image-based problem descriptions (\texttt{<image>} tags), posing additional comprehension barriers for text-only models. As shown in Figure \ref{fig:Dataset_Score}, on HumanEval, most models achieve pass@1 score above 60\%, while performance drops significantly on CodeContests (average pass@1 < 3\%).

\textbf{Parameter Size Effects}: We observe a strong positive correlation between model parameter size and code generation performance. For example, increasing the parameters from 1.5B (Yi-Coder) to 32B (Qwen2.5-Coder) yields a 59\% improvement in the LeetCodeSample pass@5 score and 56\% improvement in LeetCodeSample pass@1 score. Similarly, on CodeContests, Yi-Coder only achieved 0.61\% in pass@1 score and 1.21\% in pass@5 score, indicating that it only solved a small number of problems. In contrast, Qwen2.5-Coder obtained 4.85\% pass@1 score and 11.52\% pass@5 score, reflecting a notable improvement. After conducting a correlation analysis between parameter size and HumanEval pass@1 scores, we found that the correlation coefficient was 0.72, which exceeds 0.7. \textbf{This shows that increasing the parameter size brings about significant changes in a model performance.} However, it is worth noting that for models like the 7B CodeLlama, its average performance across the three datasets is significantly lower than that of the two 1.5B models (Yi-Coder-1.5B-Chat and Qwen2.5-Coder-1.5B-Instruct). We attribute this to continuous architectural optimizations. Newer models often have advanced architectures and training techniques, enabling them to outperform older models.

\textbf{Cost-Performance Trade-offs}: Figure \ref{fig:Cost_Pass@1_Compare} reveals a growth in computational overhead as LLM's capacity and parameter size increases. Qwen2.5-Coder-1.5B-Instruct has the highest performance as well as the largest number of parameters, and it also has the highest cost at \$1.26/M Tokens. In contrast, CodeLlama-7B-Instruct only need \$0.42/M Tokens and it has the worst performance in HumanEval with only 7B parameters, less than Qwen's parameters. Larger models need more memory requirements, with 30B+ models like Qwen2.5-Coder-32B-Instruct needing more than 60GB VRAM for BF16 inference–requiring at least four NVIDIA RTX 4090 GPUs (24GB each). Cloud costs amplify disparities: models with 16.1B+ parameters need costs that are 9× higher than those with 0-4B parameters on Fireworks (\url{https://fireworks.ai/pricing}). Although larger models achieve superior code generation performance, their operational costs become prohibitive for budget-constrained projects. \textbf{Thus, the cost-performance trade-off underscores the need to select an optimal LLM, balancing parameter scale with financial resources.}

Based on above findings, the design of a framework that can maintain model performance while effectively reducing inference costs has become a critical issue. To address this, we propose a novel framework that aims to leverage the strengths of multiple large models. The core principle of this framework is to dynamically allocate code tasks to the most suitable model based on specific task requirements. For example, simpler tasks are prioritized to smaller parameter-scale models to minimize inference costs, while complex tasks are delegated to larger parameter-scale models to ensure task completion quality. In the following section, we will provide a detailed explanation of the framework.

\section{Methodology}
\begin{figure*}[h]
	\centering
        \setlength{\abovecaptionskip}{0.1cm}
	\includegraphics[width=1\linewidth]{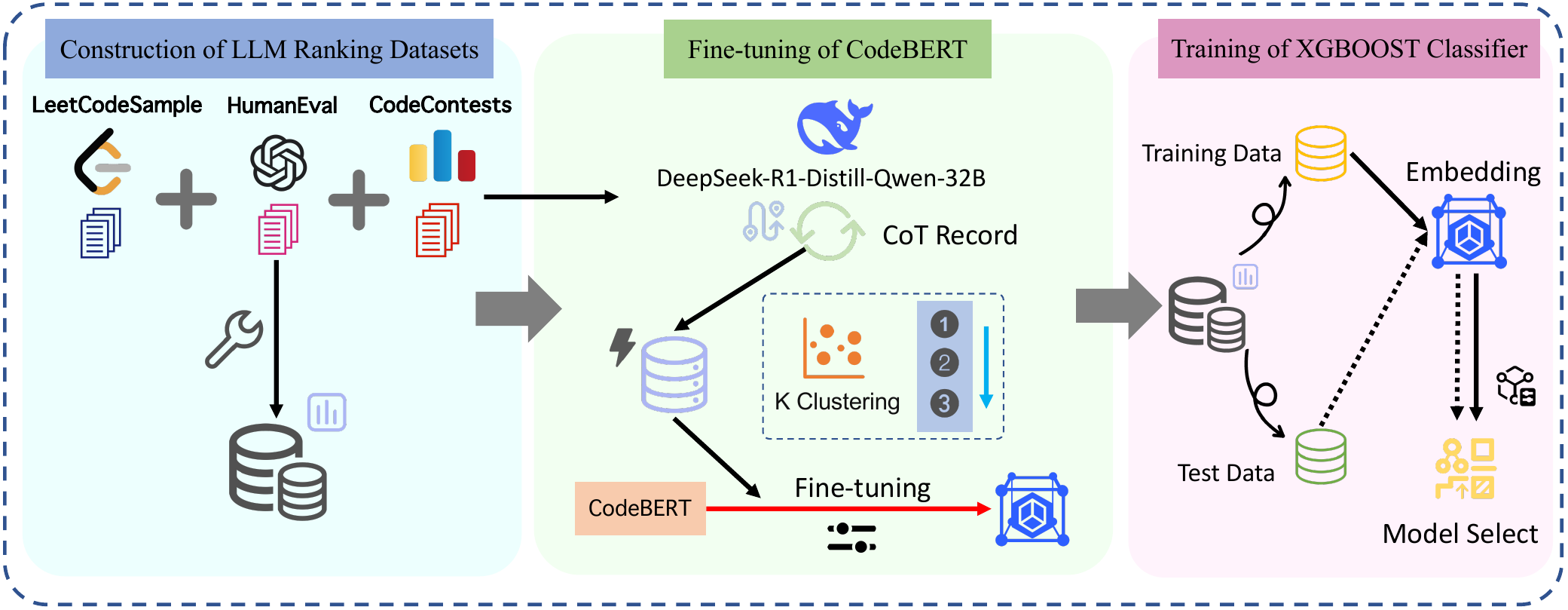}
	\caption{The overall architecture of AdaptiveLLM.}
	\label{fig:the_overall_architecture_of_AdaptiveLLM}
\end{figure*}

Figure \ref{fig:the_overall_architecture_of_AdaptiveLLM} illustrates the overall architecture of AdaptiveLLM. As a supervised learning approach, AdaptiveLLM first requires reconstructing the three datasets (Section \ref{subsec:construction_llm_ranking_datasets}) due to the absence of learning objectives in their original form. The framework consists of two core components: an embedding layer (Section \ref{subsec:fine-tuning of codebert}) and a classification layer (Section \ref{subsec:training of xgboost classifier}). Finally, we conduct a comprehensive evaluation of the AdaptiveLLM framework.

\subsection{Construction of LLM Ranking Datasets}
\label{subsec:construction_llm_ranking_datasets}
\begin{figure}[h]
	\centering
        \setlength{\abovecaptionskip}{0.1cm}
	\includegraphics[width=1\linewidth]{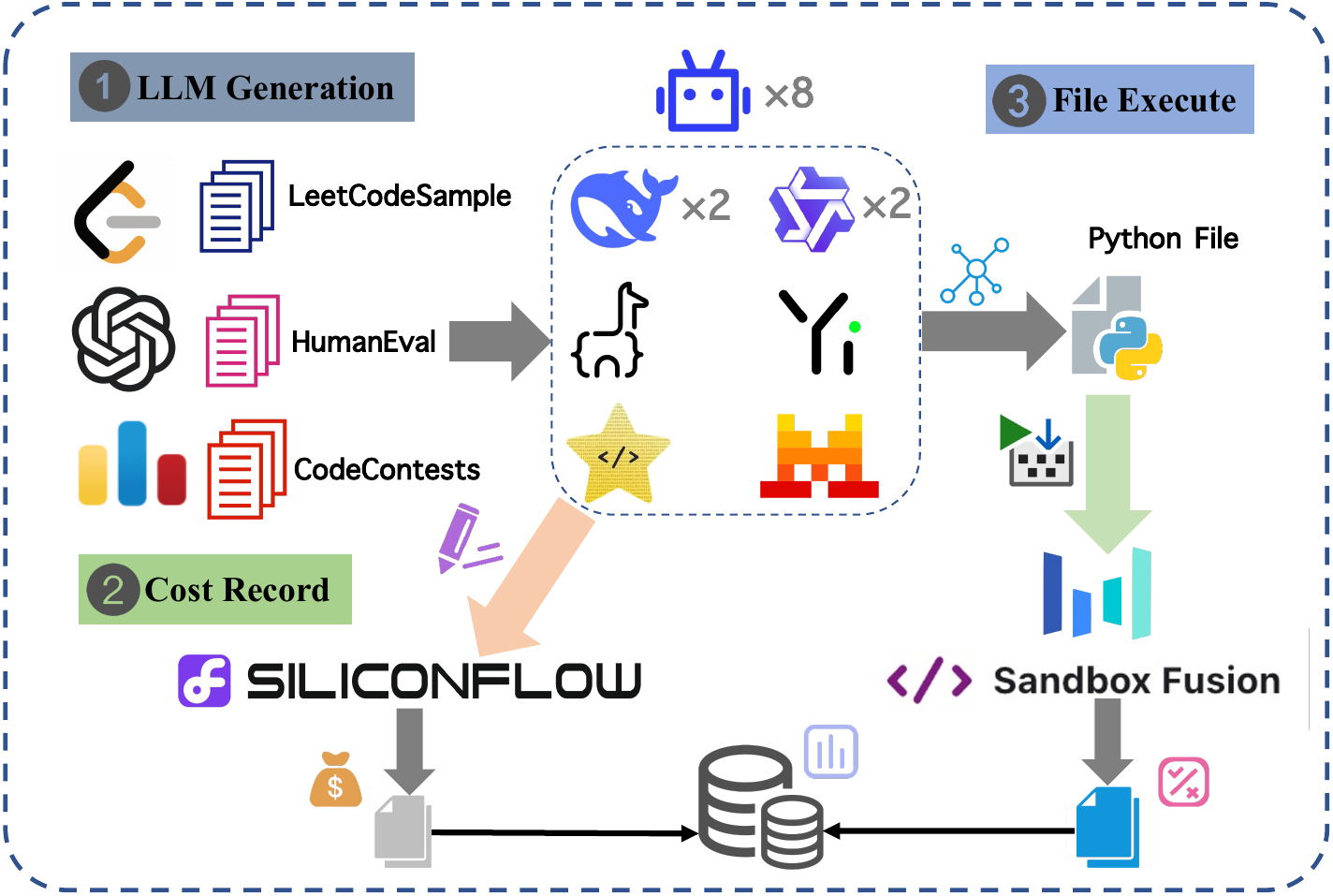}
	\caption{The process of LLM ranking dataset construction.}
	\label{fig:construction_of_LLM_ranking_datasets}
        \vspace{-0.3cm}
\end{figure}

To construct the AdaptiveLLM framework, we first conduct a research study to identify datasets and large models applied to the AdaptiveLLM framework and ranked the models used for each question based on the performance and cost of the model answers. Due to the variety of code LLMs, with their different parameters, architectures, and inference costs, to ensure that the pool of models in the AdaptiveLLM framework can cover most of the LLMs, we select eight representative LLMs among models with different parameter sizes and with different architectures. A brief description of these LLMs is provided in Section \ref{subsubsec:model-candidate-pool}. The construction process of LLM ranking dataset is shown in Figure \ref{fig:construction_of_LLM_ranking_datasets}. Constructing the LLM ranking dataset requires a total of two steps. The first step is to first have the models in the model candidate pool answer the questions in the selected dataset, and then record metrics such as correctness, token spend, and cost. The second step is to select the optimal LLM for each question based on the recorded metrics, and this step will rank the LLMs used in each question based on the metrics. Finally, the constructed dataset will be divided into training and test sets, and the problem-optimal LLM pairs in the training set will be used to train the XGBOOST classifier in Section \ref{subsec:training of xgboost classifier}.

\subsubsection{Recording of test result metrics}
\label{subsec:recording-of-test-result-metrics}
In Section \ref{sec:preliminary-study}, we conduct experiments on three datasets for eight large models. To ensure consistency in testing the generated code files, we implement the test validation in SandboxFusion sandbox environment \cite{liu2024fullstack} and record the accuracy of each answer for each model. At the same time, we also record the resource consumption metrics of the models, and the price of LLM is provided by SiliconFlow (\url{https://cloud.siliconflow.cn/models}), a cloud interfacing platform. However, given that not all of the selected models are deployed on the platform, we use the consumption data of a model with the same parameter size as a proxy for the non-deployed models. Following this step, we obtain a dataset comprising HumanEval, LeetCodeSample, and CodeContests. For each problem in the dataset, it includes five responses from each of the eight LLMs, along with their corresponding correctness results, accuracy rates, and token costs for all five responses.

\subsubsection{LLM ranking}
\label{subsubsec:LLM-ranking}
Since AdaptiveLLM operates as a supervised learning framework, for each problem, we need to predefine an optimal LLM. This optimal LLM should achieve both high accuracy in solving the problem and minimal computational costs. To accomplish this, for every problem, we rank the performance of the eight LLMs on each problem and select the top-ranked model as the optimal LLM.
So, we select LLM with the highest $\mathrm{Score}_i$ as the optimal model for each problem. $\mathrm{Score}_i$ is calculated as follows:

\begin{align}
  Score_i=\mathrm{log}(&\mathrm{max}_{j=1}^N(Tokens_j) \times MaxPrice)\times pass_i
  \notag
  \\&-\mathrm{log}(Tokens_i \times Price_i)
\end{align}

The formula reaches the optimization of model selection with double constraints. First, $pass_i$ is the test pass rate variable, which reflects the proportion of the model that passes all the test cases in five responses. Second, the resource consumption (the product of the number of Token and the inference price) is transformed into a penalty term by introducing a logarithmic operation. This makes the model's score higher the more efficient and less costly its inference is based on the guarantee of correctness of answers. Specially, $MaxPrice$ selects the highest unit price among all models as the normalized benchmark, a move that eliminates the quantitative effects arising from the differences in the pricing systems of different models.

Based on this scoring formula, we can construct a model recommendation system. For each code problem, after calculating $Score_i$ for all candidate models, the model with the highest score is selected as the optimal solution model for the problem. Due to the joint optimization of $Tokens_i$ and $Price_i$ in the formula, the selected model can satisfy the two objectives of code correctness and cost control at the same time. The result dataset covers the complete problem description, model-generated code, token consumption records, inference price details, and optimal model labels based on $Score_i$. You can find detailed information about the dataset in the public repository that we have provided.

\subsection{Fine-tuning of CodeBERT}
\label{subsec:fine-tuning of codebert}
In the preliminary study in Section \ref{sec:preliminary-study}, we find that some models perform better when dealing with more complex problems, but at the same time, these models have larger parameter sizes and consume more resources. Given this finding, we would like to incorporate the difficulty of the problem into AdaptiveLLM framework to obtain the best prediction results. However, in real-world scenarios, code problems often lack difficulty annotations. Even when such annotations are available, they may not align with the perception of difficulty as understood by LLMs \cite{ouyang2025empirical,kou2023model}. To address this challenge, we propose leveraging the length of CoT generated by inference models, such as DeepSeek R1, as a proxy for problem difficulty. This is motivated by the observation that longer CoT lengths generally correspond to higher problem complexity, providing a reliable and automated way to estimate difficulty without relying on manual annotations. So we first use DeepSeek-R1-Distill-Qwen-32B, a distilled version of the reasoning model DeepSeek R1, to assess the difficulty of each problem, and then fine-tune the CodeBERT \cite{feng2020codebert} embedding model using triplet contrast loss. 

It is important to note that we do not record cost when using DeepSeek-R1-Distill-Qwen-32B. This is because we only need to evaluate the cost of the model selected in the model candidate pool specifically when answering the question, not the cost of using the model in the training framework phase. Using DeepSeek R1 is so that the question difficulty information is embedded in the final embedding vector representation. And the involvement of DeepSeek R1 is not required when using AdaptiveLLM.

\subsubsection{Labeling problem difficulty}
\label{subsubsec:labeling-problem-difficulty}
For the constructed dataset, we will use DeepSeek-R1-Distill-Qwen-32B to record the reasoning length for each problem. As seen in Figure \ref{fig:the_process_of_difficulty_labeling}, problems of different difficulty levels show significant differences in the reasoning length. The reasoning length of a simple problem may be in a lower order of magnitude, while the reasoning length of a complex problem jumps to a higher order of magnitude. For example, the reasoning length of \texttt{CodeContests/28} problem is \num{41058}, that of \texttt{LeetCode/22} problem is \num{24806}, and that of \texttt{HumanEval/0} problem is only \num{6832}. This difference reflects the fact that when solving complex problems, the reasoning model needs more complex reasoning steps and logical processes to reach a solution. In contrast, the answer lengths remain largely consistent across problems of varying difficulty. We use DeepSeek-R1-Distill-Qwen-32B and count the average of the reasoning lengths and answer lengths on each dataset. The detailed data is shown in Table \ref{tab:Comparison of DeepSeek-R1-Distill-Qwen-32B's averages}. This implies that the size of the content output by the model is relatively stable when generating the final code answer, despite the different complexity of the questions.

\begin{table}[h]
    \centering
    \setlength{\abovecaptionskip}{0.1cm}
    \caption{Comparison of DeepSeek-R1-Distill-Qwen-32B's averages of reasoning lengths and answer lengths on different datasets shows a large variability in reasoning lengths, while answer lengths are not discriminatory.}
    \begin{tabular}{l|ccc}
    \toprule
         \textbf{Avg Length} & \textbf{HumanEval} & \textbf{LeetCode} & \textbf{CodeContests} \\
         \midrule
         Reasoning Len & 7837.46 & 24606.85 & 40882.77\\
         Answering Len & 1985.59 & 2676.79 & 3192.45\\
    \bottomrule
    \end{tabular}
    \label{tab:Comparison of DeepSeek-R1-Distill-Qwen-32B's averages}
    \vspace{-0.3cm}
\end{table}

Considering this, we decided to use the k-means clustering algorithm to cluster and analyze the reasoning lengths of all questions. Following the clustering process, three distinct difficulty levels, labeled as 1, 2, and 3, were identified. That is, we will let the CoT lengths recorded by DeepSeek R1 be used as inputs in all problems, and using the k-means algorithm, we will classify the problems into 3 categories representing three different levels of difficulty: easy, medium, and hard. And then, these classification results will be incorporated into the subsequent CodeBERT fine-tuning process.

\begin{figure}[h]
	\centering
        \setlength{\abovecaptionskip}{0.1cm}
	\includegraphics[width=1\linewidth]{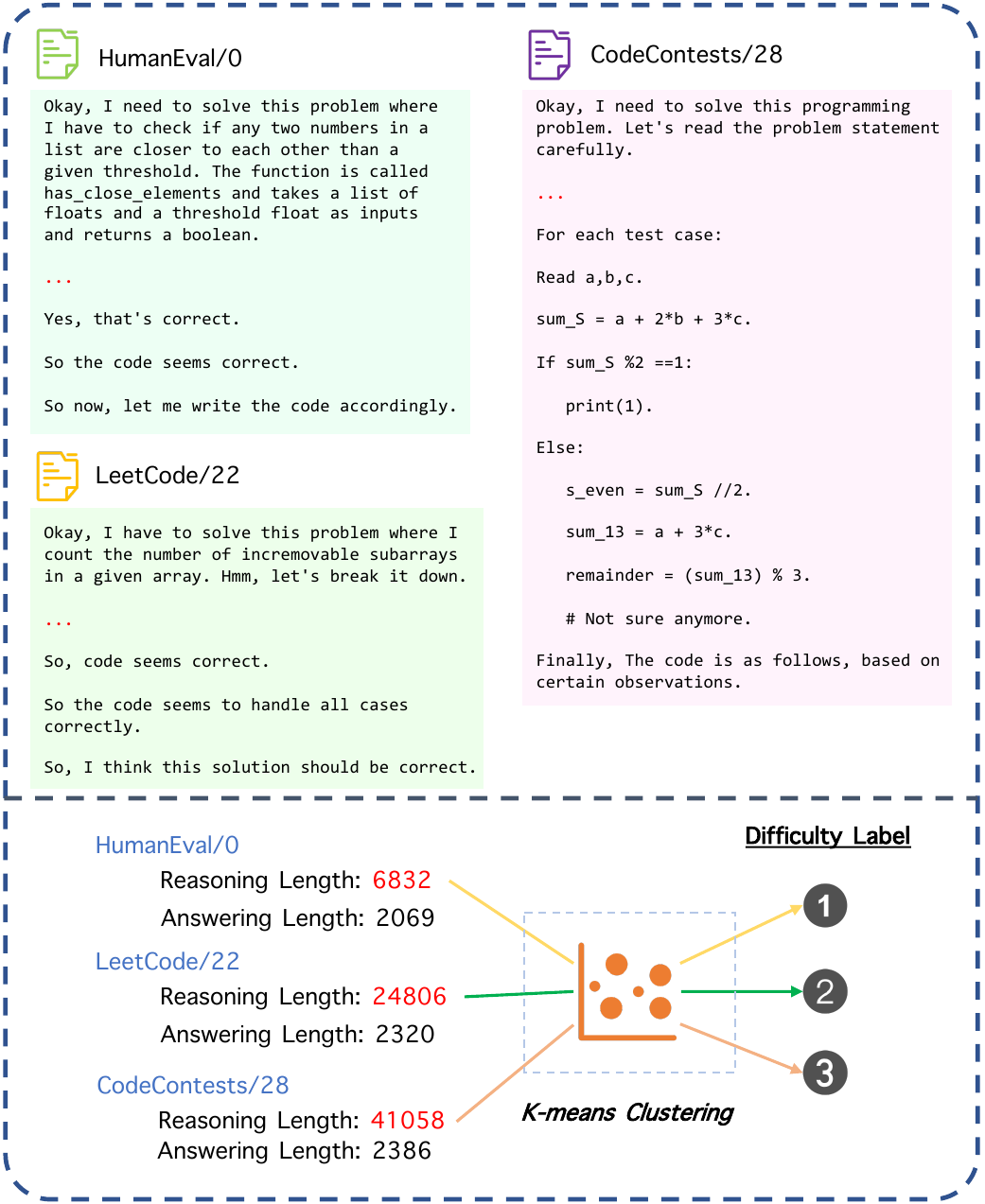}
	\caption{The process of CoT difficulty labeling.}
	\label{fig:the_process_of_difficulty_labeling}
        \vspace{-0.3cm}
\end{figure}

\subsubsection{Contrast loss learning}
The purpose of contrast learning is to allow the CodeBERT embedder to learn the difficulty information of the problem. After contrast learning process, the problem is transformed into an embedding vector that will have difficulty information, which is useful for our subsequent LLM selection.

In contrast learning, optimization process is driven by triplet loss, where each triplet comprises an anchor sample, a positive sample, and a negative sample \cite{hoffer2015deep,wei2022clear}. Specifically, anchored samples are problems that require optimization of the embedding space, positive samples are semantically of the same difficulty as anchored samples, and negative samples are semantically of a different difficulty than anchored and positive samples. At this point three clusters($C_1, C_2, C_3$) have been obtained after labeling, representing the three different difficulties of the problem. For each anchored problem $x_i$, the following sampling method is used to construct pairs of positive and negative samples:
\begin{itemize}[left=0pt]
\item \textit{Positive sample}: Randomly select a problem with a similar CoT length from the difficulty cluster $C_i$ that is the same as the anchored sample.
\item \textit{Negative sample}: Randomly choose one of the two clusters $C_j$ and $C_k$ with different difficulties, and then randomly select a problem with a different CoT length from this cluster.
\end{itemize}

Then, we can use triplet data to fine-tune the CodeBERT embedding model. The core objective of contrastive learning is to bring queries with similar features and complexities closer together while pushing unrelated queries with different complexities further apart \cite{oord2018representation}. The design of triplet data enables the model to learn semantic representations by associating the anchor query with its positive counterparts, positioning them closer in the embedding vector space. On the contrary, we expect the model to push unrelated queries farther away from the anchor query.

Specifically, first, we use CodeBERT to extract the embedding representations of each word in every query. These word-level embeddings are aggregated through a pooling layer to generate fixed-size sentence-level embeddings. We denote the mapping of problem $x_i$ into the embedding space $\mathbb{R}^{p}$ by CodeBERT before fine-tuning as $\mathcal{E}(x_i;\mathbf{w})$. Based on this, we define the following contrastive loss formula:

\begin{align}
Loss&=\max\left(0,\frac{\mathbf{a}\cdot\mathbf{n}}{\|\mathbf{a}\|\|\mathbf{n}\|}-\frac{\mathbf{a}\cdot\mathbf{p}}{\|\mathbf{a}\|\|\mathbf{p}\|}+margin\right)
\notag
\\&=\max\left(0,\mathbf{a}\cdot\mathbf{n}-\mathbf{a}\cdot\mathbf{p}+margin\right)
\end{align}

\begin{itemize}[left=0pt]
    \item $\mathbf{a}=\mathcal{E}(x_{\text{anchor}};\mathbf{w})$: The embedding of the anchor sample.
    \item $\mathbf{p}=\mathcal{E}(x_{\text{positive}};\mathbf{w})$: The embedding of the positive sample.
    \item $\mathbf{n}=\mathcal{E}(x_{\text{negative}};\mathbf{w})$: The embedding of the negative sample.
\end{itemize}

The goal of this loss formula is to learn an embedding space where semantically similar sentences are close to each other, while semantically different sentences are far from each other. The $margin$ in the formula is a positive value (by default set to 1), which is used to define the minimum gap between the distance between the anchor and the positive sample and the distance between the anchor and the negative sample. This parameter ensures that negative samples are not just simply pushed away from positive samples but maintain a meaningful distance. In addition, $\max(0,\cdot)$ ensures that the loss value is non-negative. This means that when the distance between the negative sample and the anchor is already large enough, the loss will be zero, indicating that the current triplet does not need further update. 

\subsection{Training of XGBOOST Classifier}
\label{subsec:training of xgboost classifier}
Once CodeBERT is fine-tuned, it will serve as a feature extractor for the embedding layer, assisting the XGBoost classifier \cite{chen2016xgboost} in supervised learning using the LLM ranking dataset constructed in Section \ref{subsec:construction_llm_ranking_datasets}. 

The rationale for selecting XGBoost as the final classifier is evident. Our objective is to construct a framework for selecting the optimal code LLM that ensures both accuracy and minimal resource consumption. Directly incorporating a neural network (NN) into CodeBERT's output layer would contradict our goal due to its substantial computational demands. Therefore, we opted for the lightweight classifier architecture XGBoost, which has been validated as effective in prior research \cite{aydin2021performance,chen2018xgboost}.

In our framework, the input features are the embedding vectors generated by the fine-tuned CodeBERT, and the target labels are the optimal model labels corresponding to each problem (specifically, the model that attains the highest score calculated by $\mathrm{Score}_i$ in Section \ref{subsubsec:LLM-ranking}). It should be noted that during the training process, the classifier never has access to the test set to avoid data leakage.

\section{Experimental Setups}

\subsection{Studied Models}
\begin{table*}[h]
    \centering
    \setlength{\abovecaptionskip}{0.1cm}
    \caption{Detailed information of LLMs used in our experiments. We do not report the API price for Reasoning Models because these models are used to assess the difficulty of coding problems through CoT length rather than being part of the candidate models for selection.}
    \resizebox{\linewidth}{!}{ 
    \begin{tabular}{l|lccc}
    \toprule
         \textbf{Type} & \textbf{Model}&  \textbf{Size} & \textbf{Model Link} & \textbf{API Price(\$/M tokens)}\\
         \midrule
         \multirow{8}{1.5cm}{Candidate Pool} 
         & \textbf{Yi-Coder-1.5B} & 1.5B & \url{https://hf.co/01-ai/Yi-Coder-1.5B-Chat} & 0.14 \\ 
         & \textbf{Qwen2.5-Coder-1.5B-Instruct} & 1.5B & \url{https://hf.co/Qwen/Qwen2.5-Coder-1.5B-Instruct} & 0.14 \\ 
         & \textbf{CodeLlama-7B-Instruct} & 7B & \url{https://hf.co/meta-llama/CodeLlama-7b-Instruct-hf} & 0.42 \\ 
         & \textbf{Starcoder2-15B-Instruct} & 15B & \url{https://hf.co/bigcode/starcoder2-15b-instruct-v0.1} & 0.72 \\ 
         & \textbf{DeepSeek-Coder-V2-Lite-Instruct} & 16B & \url{https://hf.co/deepseek-ai/DeepSeek-Coder-V2-Lite-instruct} & 0.72 \\
         & \textbf{Codestral-22B} & 22B & \url{https://hf.co/mistralai/Codestral-22B-v0.1} & 0.95 \\
         & \textbf{DeepSeek-Coder-33B-Instruct} & 33B & \url{https://hf.co/deepseek-ai/deepseek-coder-33b-instruct} & 1.26 \\
         & \textbf{Qwen2.5-Coder-32B-Instruct} & 32B & \url{https://hf.co/Qwen/Qwen2.5-Coder-32B-Instruct} & 1.26 \\
         \midrule
         \multirow{4}{1.5cm}{Reasoning Models} 
         & DeepSeek-R1-Distill-Qwen-1.5B & 1.5B & \url{https://hf.co/deepseek-ai/DeepSeek-R1-Distill-Qwen-1.5B} & - \\
         & DeepSeek-R1-Distill-Qwen-7B & 7B & \url{https://hf.co/deepseek-ai/DeepSeek-R1-Distill-Qwen-7B} & - \\
         & DeepSeek-R1-Distill-Qwen-14B & 14B & \url{https://hf.co/deepseek-ai/DeepSeek-R1-Distill-Qwen-14B} & - \\
         & DeepSeek-R1-Distill-Qwen-32B & 32B & \url{https://hf.co/deepseek-ai/DeepSeek-R1-Distill-Qwen-32B} & - \\
    \bottomrule
    \end{tabular}
    }
    \label{tab:the-details-of-LLMs-used-in-our-experiments}
    \vspace{-0.3cm}
\end{table*}

\subsubsection{Model candidate pool}
\label{subsubsec:model-candidate-pool}
Table \ref{tab:the-details-of-LLMs-used-in-our-experiments} shows the code models used in our framework. To ensure that the model pool can cover models with different parameter sizes, we select multiple models ranging from a minimum of 1.5 billion parameters to a maximum of 33 billion parameters. This selection aims to provide suitable code models in response to problems of varying difficulties. The inference prices of the large language models listed in the table are sourced from the SiliconCloud cloud inference platform (\url{https://www.siliconflow.com/en/pricing}). However, since some models (such as Starcoder2-15B-Instruct and CodeLlama-7B-Instruct) have not been listed on this platform yet, for these models that are not covered, we assume that their costs are close to those of models with similar parameter sizes. Therefore, we use the prices of models with similar parameter sizes as alternative estimates. For code LLM using, we use RTX 4090 locally to deploy these models. As far as we know, the NVIDIA RTX 4090 GPU offers only 24GB of VRAM, necessitating multi-GPU deployment for models with 15B parameters or larger. Since inference speed is not a consideration in our evaluation framework and cost metrics are standardized through cloud service provider platforms, the use of single vs. multi-GPU configurations does not affect the experimental outcomes.
Considering that there are differences in the default parameter settings of different code models, to ensure the consistency of the experiments, we standardize the model settings: for models with their own default parameters, the parameter settings recommended by the official are adopted; for models without provided default parameters, we uniformly set $do\_sample = True$, $temperature = 0.3$, $top\_p = 0.95$, and $top\_k = 20$. Additionally, all models are implemented using the Hugging Face Transformers library in Python for loading LLMs.

\subsubsection{Reasoning model used in experiments}
Table \ref{tab:the-details-of-LLMs-used-in-our-experiments} shows the reasoning models used in our experiments. During the construction of AdaptiveLLM, we employ DeepSeek-R1-Distill-Qwen-32B to measure and record the CoT length. To validate the rationale of using CoT length as a proxy for problem complexity, we conduct experiments in RQ1 in Section \ref{subsec: rq1} using multiple DeepSeek R1 variants with different parameter scales. For models with relatively smaller parameter sizes, such as DeepSeek-R1-Distill-Qwen-1.5B and DeepSeek-R1-Distill-Qwen-7B, a single RTX 4090 GPU is deployed for inference. Conversely, models with larger parameter sizes, including DeepSeek-R1-Distill-Qwen-14B and DeepSeek-R1-Distill-Qwen-32B, are used via API, with the API service also provided by SiliconCloud (\url{https://api.siliconflow.cn/v1}). Although we utilized both local and cloud service deployment methods for DeepSeek, we assume that their impact on the final CoT length is negligible. This is because the difference in deployment only affects the speed of inferencing, not the quality of the answer.

\subsection{Metrics and Baselines}
\subsubsection{Metrics}
\begin{itemize}[left=0pt]
    \item \textbf{pass@k\_score}. This metric measures the probability that at least one correct solution is generated among the top K answers. It is commonly used to evaluate the performance of code generation models. A higher pass@k indicates a better ability of the model to generate correct code within the top K responses.
    \item \textbf{pass@k\_token}. This metric represents the average number of tokens consumed per problem during the pass@k test. Tokens are the basic units of input and output in the model, and the quantity of tokens used directly reflects the computational resources required for the test. 
    \item \textbf{pass@k\_price}. This metric indicates the average cost per problem incurred during the pass@k test and is calculated by multiplying the pass@k\_token by the inference price per token. Different models may have different pricing for their tokens, meaning that the same number of tokens could result in significantly different costs depending on the model. By using pass@k\_price, we can more accurately reflect the true resource cost of running the pass@k test for each model.
\end{itemize}

\subsubsection{Baselines}
In our experiments, we choose ComplexityNet \cite{bae2023complexitynet} as the baseline method for comparison. Its core mechanism involves pre-interactions between LLMs and tasks to determine the complexity of the tasks. Specifically, the framework allows LLMs with varying capabilities (Code Llama, GPT-3.5, and GPT-4) to attempt solving the tasks multiple times. Based on the correctness of the model outputs, tasks are assigned difficulty labels. These labels are then used by a smaller model, DaVinci-002 \cite{brown2020language}, to classify and allocate tasks to the most suitable models.

Given the potential unavailability of certain LLMs, we have implemented substitutions in our experimental setup. Specifically, we have employed gpt-4o-2024-11-20 as an alternative for GPT-4 and Qwen2.5-7B-Instruct as an alternative to DaVinci-002. These alternative models are chosen to closely approximate the performance of the models that we used in AdaptiveLLM, thereby ensuring the reliability, fairness and validity of the experimental results.

\subsection{Implementation Details}
In Section \ref{subsec:fine-tuning of codebert}, during the fine-tuning of CodeBERT, we employed the AutoTokenizer and AutoModel classes from the Hugging Face Transformers library. And for encoder-only architectures like BERT, the maximum input sequence length is 512. Consequently, we applied truncation to inputs exceeding this limit, with implementation details documented in our repository. This strategy is rational. Because problem descriptions typically appear at the beginning of input sequences, ensuring that the essential semantic information remains preserved in generated embedding vectors.

In Section \ref{subsubsec:labeling-problem-difficulty}, we log the optimal CoT length for reasoning models. Given the variability in the lengths of CoT, we employ the DeepSeek-R1-Distill-Qwen-32B model to generate ten responses for each problem. The median length of these responses is selected as the optimal CoT length for each problem. This approach ensures that the chosen length reflects the complexity of the problem. The same methodology is applied in Section \ref{subsec: rq1} to get CoT lengths for reasoning models with varying parameter sizes.

We cap the maximum number of tokens generated by reasoning models at \num{16384}. This setting allows most problems to generate complete CoT. However, for exceptionally complex problems where the CoT length exceeds this limit, we record the length as 0 and exclude these instances from our dataset. For code models in model candidate pool, we set the maximum number of tokens to \num{2048}, ensuring complete responses for all problems without truncation.

In Section \ref{subsec:training of xgboost classifier}, during XGBoost training process, the dataset is randomly divided into training and testing subsets at a ratio of 70\% to 30\%. Specifically, the training set consists of 411 coding problems, while the testing set includes 178 coding problems. For the implementation of XGBoost, we utilized the XGBClassifier class from the xgboost library to train the XGBoost model. Detailed parameter configurations are documented in our repository.

\section{Experimental Results and Analysis}
To assess the effectiveness of AdaptiveLLM, we conduct a thorough evaluation addressing the following research questions:
\begin{itemize}[left=0pt]
    \item \textbf{RQ1: Rationality Analysis} - How does the difficulty assessed by the CoT length generated by reasoning models compare with the true difficulty labels? Is it reasonable? Is it affected by the parameter size of the reasoning models?
    \item \textbf{RQ2: Overall Performance} - How does AdaptiveLLM perform on the test set metrics compared to a single LLM and baseline method?
    \item \textbf{RQ3: Ablation Study} - What is the impact of the embedding layer fine-tuning in AdaptiveLLM on the overall performance?
\end{itemize}

\subsection{RQ1: Rationality Analysis}
\label{subsec: rq1}
\subsubsection{RQ1-1: Is there a significant difference between the difficulty assessed by CoT length and the true difficulty labels?}
In the LeetCodeSample and CodeContests datasets, difficulty labels are assigned based on human performance. Specifically, the difficulty in LeetCodeSample is determined by the acceptance rate (acRate), and CodeContests relies on competition ratings (cf\_rating). These metrics reflect the actual performance of human solvers in terms of success rates.

To enable comparative analysis, we employ the K-means clustering algorithm on each dataset to classify problems with actual difficulty labels into two categories: easy and hard. We then apply the same binary classification to the difficulty labels generated by the DeepSeek-R1-Distill-Qwen-32B model, which are based on the CoT length. Through this process, we compare the model's predicted difficulty against the true difficulty and construct a confusion matrix to analyze.

\begin{figure}[h]
	\centering
        \setlength{\abovecaptionskip}{0.1cm}
	\includegraphics[width=1\linewidth]{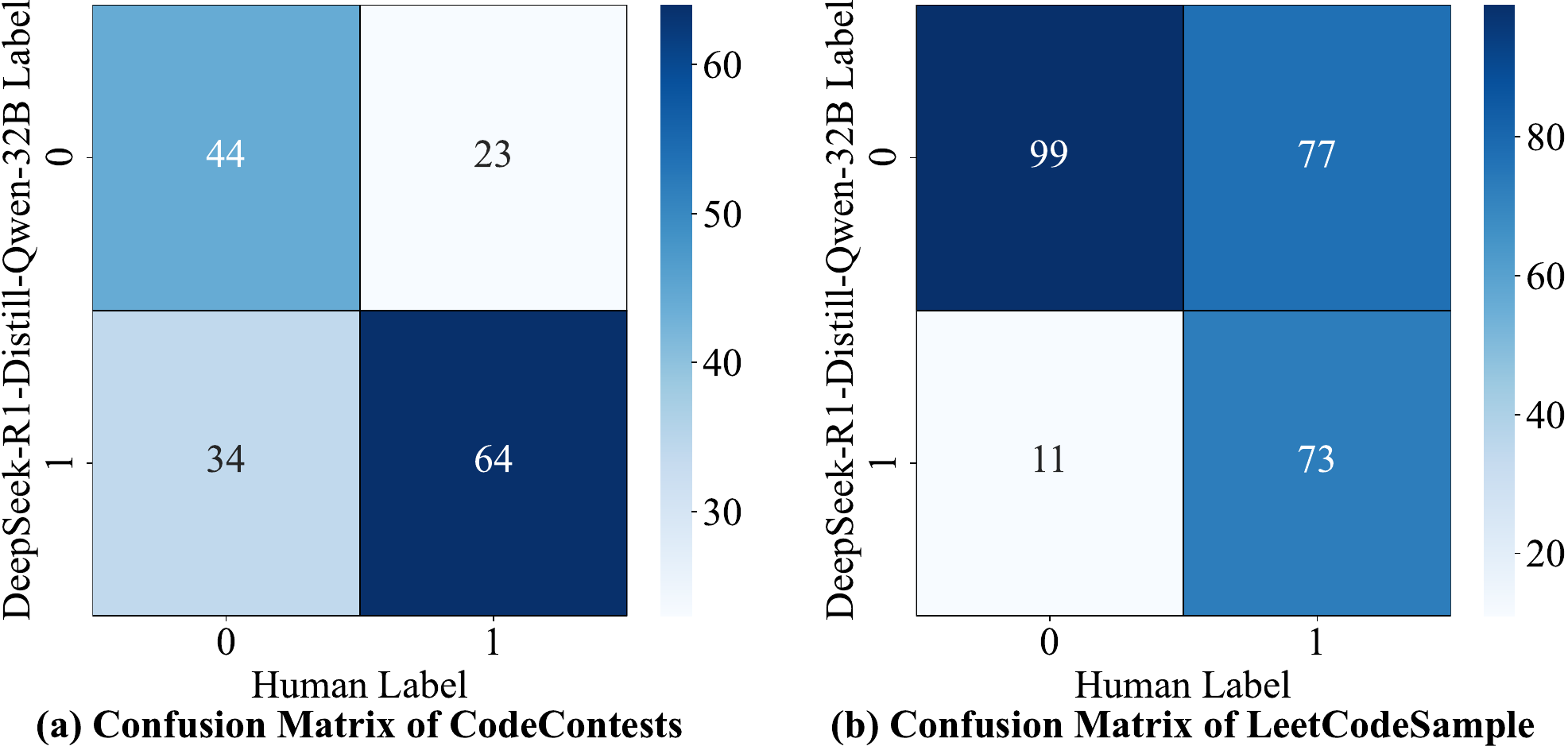}
	\caption{The confusion matrix between CoT difficulty and true difficulty.}
	\label{fig:the-confusion-matrix-with-true-labels}
        \vspace{-0.3cm}
\end{figure}

As shown in Figure \ref{fig:the-confusion-matrix-with-true-labels}, even under the binary clustering, there is a significant difference between the difficulty assessed by CoT length and the true difficulty labels. Specifically, in the CodeContests dataset, a total of 57 problems were classified into different difficulty categories by the two methods, while in the LeetCodeSample dataset, this number reached 88. Notably, in the LeetCodeSample dataset, 77 problems labeled as "hard" (label 1) by the true labels were classified as "easy" (label 0) by the CoT length assessment. This finding indicates a substantial difference between the difficulty labels assigned by the R1 model based on CoT length and the original difficulty labels in the dataset. 

The original difficulty labels were created by humans, but the actual problem-solving is done by LLMs. If we use the difficulty with which humans view a question to guide LLM in answering the question, it may be biased. So, is LLM's perception of the difficulty of the question is more in line with LLM's perspective of answering the question?

\subsubsection{RQ1-2: Which difficulty assessment method is more practical for LLM?}
Given the inconsistency in difficulty assessment, we design and conduct the following experiment. After logging the CoT lengths, we normalized the extracted CoT lengths using a normal distribution normalization to ensure comparability of the data. Meanwhile, each problem in the LeetCodeSample dataset is accompanied by a difficulty label based on human perception, which is defined as $1 - acRate$. Problems with higher difficulty have lower acceptance rates, resulting in higher difficulty scores. To facilitate comparison with the CoT length, we also normalize these human perception-based difficulty labels using the same normal distribution normalization. Finally, we extract problems from LeetCodeSample that could be correctly answered by Qwen2.5-Coder-32B-Instruct and creat box plot for these problems based on the normalized difficulty data from both methods.

\begin{figure}[h]
	\centering
        \setlength{\abovecaptionskip}{0.1cm}
	\includegraphics[width=1\linewidth]{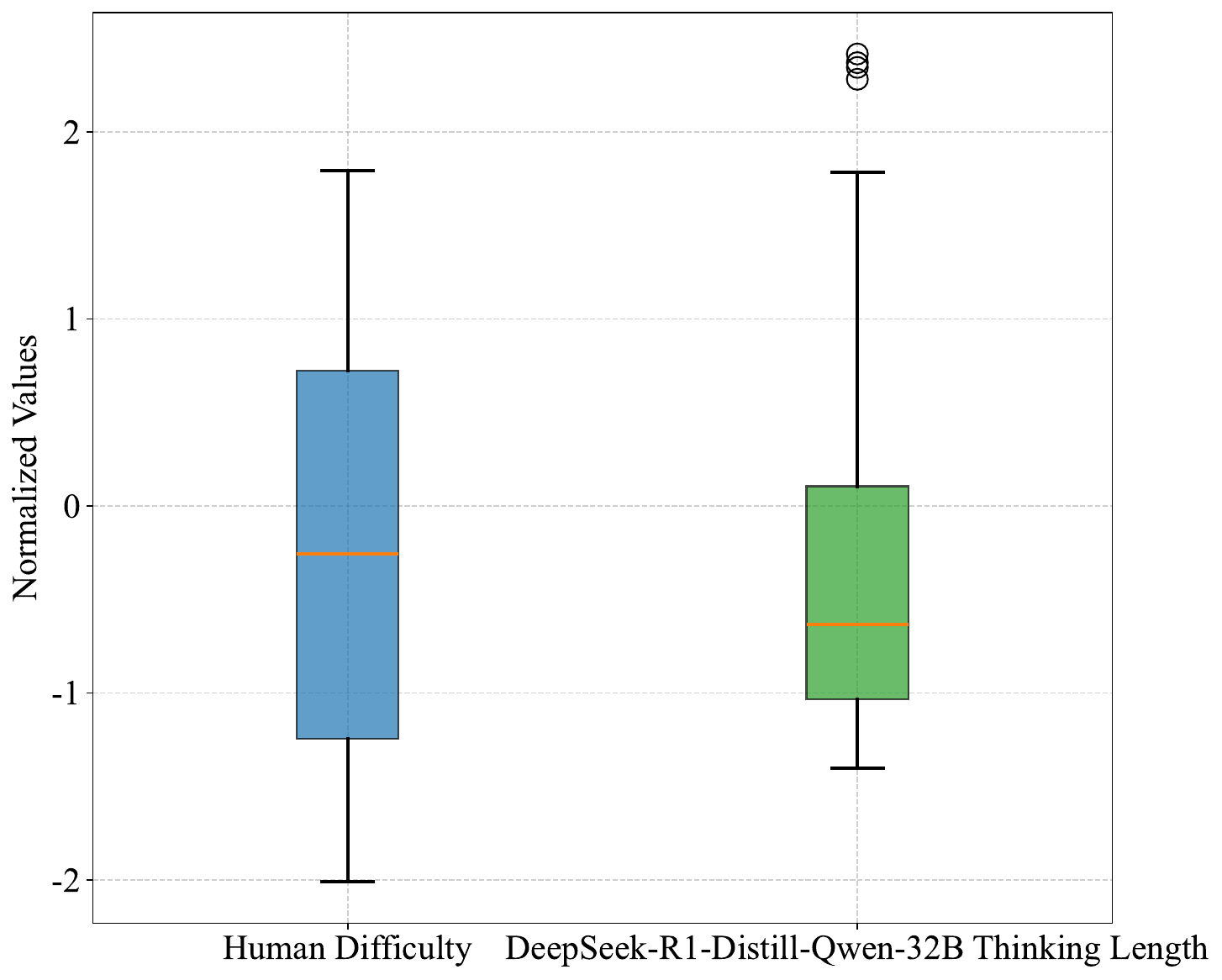}
	\caption{The box plot for comparison of CoT difficulty and true difficulty.}
	\label{fig:the-box-plot-for-comparison-of-difficulty}
        \vspace{-0.3cm}
\end{figure}

The box plot is shown in Figure \ref{fig:the-box-plot-for-comparison-of-difficulty}. Given that the selected data are from problems that the large model can answer correctly, it is expected that the difficulty labels of these problems would generally be low. Meanwhile, the corresponding CoT lengths of the reasoning model would be relatively short. As observed from the box plot, approximately 75\% of the problems have CoT lengths less than 0. This indicates that the majority of the difficulty labels based on the reasoning model's CoT lengths are relatively simple, which aligns with our selection of problems that LLMs can correctly answer. However, for the human labeled data, many scores are above 0, indicating that the actual programmers still consider these selected problems to have a certain level of difficulty. This is inconsistent with the performance of LLMs on the questions, as LLMs answer all of these questions correctly.

Moreover, the overall distribution of the difficulty labels based on CoT length is more concentrated compared to those based on human perception. This implies that CoT length tends to reflect that these promblems are generally simple. However, there are some outliers in the CoT length data. These outliers exhibit significantly longer CoT lengths that deviate from the majority of the distribution. This observation underscores that CoT still demonstrates a degree of uncertainty, potentially overcomplicating some otherwise simple problems. But it is still reasonable to apply CoT length for the few cases which can be ignored.

The results overall indicate that using CoT length as a criterion for classifying problem difficulty is rational. More importantly, compared to the human difficulty labels, it may more accurately reflect the actual performance of LLMs in code generation tasks. For example, certain logical reasoning problems that are complex for humans may be relatively straightforward for LLMs, while some challenges that are difficult for LLMs might be easy for humans. Such phenomena are not uncommon in the reasoning processes of LLMs. We believe this finding provides valuable insights and opens new perspectives for future research.

\subsubsection{RQ1-3: What are the differences in CoT lengths generated by reasoning models of different parameter sizes?}
\begin{table}[h]
    \centering
    \setlength{\abovecaptionskip}{0.1cm}
    \caption{The comparison of CoT length generated by models with different parameter sizes.}
    \begin{tabular}{l|cc}
    \toprule
         \textbf{Model Group} & \textbf{ARI} & \textbf{FMI} \\
         \midrule
         1.5B - 14B & 0.4650 & 0.6538 \\
         7B - 14B & 0.6540\ua{0.189} & 0.7778\ua{0.124} \\
         \midrule
         1.5B - 32B & 0.3476 & 0.5845 \\
         7B - 32B & 0.4845\ua{0.1369} & 0.6741\ua{0.0896} \\
         14B - 32B & 0.5711\ua{0.2235} & 0.7271\ua{0.1426} \\
    \bottomrule
    \end{tabular}
    \label{tab:comparison-of-CoT-clustering}
    \vspace{-0.3cm}
\end{table}

\begin{table*}[h]
    \centering
    \setlength{\abovecaptionskip}{0.1cm}
    \caption{Performance comparison between AdaptiveLLM and baselines.}
    \begin{tabular}{l|ccc|ccc}
    \toprule
    \multirow{2.5}{*}{\textbf{Model}} & \multicolumn{3}{c|}{\textbf{pass@1}} & \multicolumn{3}{c}{\textbf{pass@5}} \\
    \cmidrule(lr){2-4} \cmidrule(lr){5-7}
    & \textbf{score} & \textbf{token} & \textbf{price(\$)} & \textbf{score} & \textbf{token} & \textbf{price(\$)}\\
    \midrule
    Yi-Coder-1.5B & 26.41\% & 329.02 & 4.61e-05 & 35.96\% & 1659.47 & 23.23e-05 \\
    Qwen2.5-Coder-1.5B-Instruct & 29.78\% & 390.71 & 5.47e-05 & 39.33\% & 1904.76 & 26.67e-05 \\
    CodeLlama-7B-Instruct & 17.98\% & 522.68 & 21.95e-05 & 29.21\% & 2472.89 & 103.86e-05 \\
    Starcoder2-15B-Instruct & 29.21\% & 193.47 & 13.93e-05 & 39.89\% & 957.97 & 68.97e-05 \\
    DeepSeek-Coder-V2-Lite-Instruct & 49.44\% & 563.68 & 40.58e-05 & 58.43\% & 2837.76 & 204.32e-05 \\
    Codestral-22B & 44.38\% & 476.48 & 45.27e-05 & 52.81\% & 2383.47 & 226.43e-05 \\
    DeepSeek-Coder-33B-Instruct & 39.89\% & 359.61 & 45.31e-05 & 49.44\% & 1839.34 & 231.76e-05 \\
    Qwen2.5-Coder-32B-Instruct & 57.87\% & 763.57 & 96.21e-05 & 71.35\% & 3834.90 & 483.20e-05 \\
    \midrule
    GPT3.5 & 35.39\% & 130.42 & 19.56e-05 & 45.51\% & 646.28 & 96.94e-05 \\
    GPT4o & 60.67\% & 551.62 & 551.62e-05 & 72.47\% & 2741.79 & 2741.79e-05 \\
    \midrule
    ComplexityNet & 37.08\% & 380.30 & 256.83e-05 & 50.56\% & 1951.39 & 1306.66e-05 \\
    \textbf{AdaptiveLLM} & \textbf{44.94\%}\ua{7.86} & 428.80\ua{48.5} & \textbf{28.49e-05}\da{228.34} & \textbf{56.18\%}\ua{5.62} & 2094.98\ua{143.59} & \textbf{140.07e-05}\da{1166.59} \\
    AdaptiveLLM (w/o finetune) & 43.82\%\da{1.12} & 411.84\da{16.96} & 26.74e-05\da{1.75} & 55.62\%\da{0.56} & 2037.09\da{57.89} & 133.99e-05\da{6.08} \\
    \bottomrule
    \end{tabular}
    \label{tab:the-performance-of-AdaptiveLLM}
    \vspace{-0.3cm}
\end{table*}

We further investigate whether the size of model parameters affects the CoT length and its corresponding difficulty classification. To address this question, we select DeepSeek-R1-Distill-Qwen models with varying parameter sizes (1.5B, 7B, 14B, and 32B), extract the CoT lengths generated by each model, and conduct clustering analysis on these data. Subsequently, we compare the difficulty classifications of different parameter size models in a pairwise manner.

The results reveal that the consistency of difficulty classification significantly increases as the model parameter size grows. In the clustering comparison among the 1.5B, 7B, 14B, and 32B reasoning models, it is observed that as the parameter size increases, their classification results gradually converge with those of the 32B R1 reasoning model. This trend is quantified using the Adjusted Rand Index (ARI) and Fowlkes-Mallows Index (FMI), as shown in Table \ref{tab:comparison-of-CoT-clustering}. These metrics measure the similarity between the clustering results of models with different parameter sizes. Results show that model group with closer parameter sizes achieve higher ARI and FMI values, indicating stronger consistency in difficulty classification. For example, the ARI and FMI values between the 14B and 32B models are 0.5711 and 0.7271, respectively, while the corresponding values between the 1.5B and 32B models are only 0.3476 and 0.5845.

This finding suggests that smaller models exhibit randomness in CoT length, leading to less stable classifications. In contrast, larger models show a stronger correlation between CoT length and actual difficulty, improving classification stability and reliability.

\vspace{1mm}
\begin{custommdframed}
\textit{Answer to RQ1:} The difficulty assessment based on CoT length of reasoning models differs from human. Nevertheless, the CoT length method seems to better reflect the actual performance of LLMs when solving problems. And reasoning models with larger parameter sizes tend to generate more stable and consistent CoT.
\end{custommdframed}
\vspace{1mm}

\subsection{RQ2: Overall Performance}
\subsubsection{RQ2-1: How does the overall performance of AdativeLLM compare with single LLM?}
Table \ref{tab:the-performance-of-AdaptiveLLM} presents a detailed comparison of key metrics, including pass@1 and pass@5 score, token consumption, and inference cost, for eight code LLMs and the AdaptiveLLM framework on the benchmark dataset.

Results demonstrate that AdaptiveLLM framework exhibits significant performance advantages. In terms of accuracy, the AdaptiveLLM framework achieves pass@1 score of 44.94\% and pass@5 score of 56.18\%. These results not only surpass those of the Codestral-22B model with 22B parameters (pass@1 score of 44.38\% and pass@5 score of 55.28\%), but also show a clear superiority compared to the larger-parameter DeepSeek-Coder-33B-Instruct model (pass@1 score of 39.89\% and pass@5 score of 49.44\%). Notably, while maintaining a high level of accuracy, the token consumption of the AdaptiveLLM framework (pass@1: 428.80 and pass@5: 2094.98) is significantly lower than that of most larger-parameter models, such as Qwen2.5-Coder-32B-Instruct (pass@1: 763.57 and pass@5: 3834.90).

In terms of cost-effectiveness, AdaptiveLLM framework demonstrates remarkable economic efficiency. Its inference cost (pass@1: 28.49e-05\$ and pass@5: 140.07e-05\$) is comparable to CodeLlama-7B-Instruct model (pass@1: 21.95e-05\$ and pass@5: 103.86e-05\$), while achieving significantly higher accuracy. Compared to larger-parameter models, the cost control advantage of the AdaptiveLLM framework is even more evident. For example, compared to the Codestral-22B model (pass@1: 45.27e-05\$ and pass@5: 226.43e-05\$), the AdaptiveLLM framework reduces inference cost by approximately 37\% while maintaining a certain level of accuracy.

\subsubsection{RQ2-2: How does the overall performance of AdativeLLM compare with sota baseline method?}
To comprehensively evaluate AdaptiveLLM against the state-of-the-art baseline method ComplexityNet \cite{bae2023complexitynet}, we analyze both accuracy and cost-efficiency metrics. AdaptiveLLM achieves a pass@1 score of 44.94\%, outperforming ComplexityNet by 7.86\%, while consuming moderately more tokens (428.80 vs. 380.30). This accuracy gain is maintained in pass@5, where AdaptiveLLM scores 56.18\%, surpassing ComplexityNet by 5.62\%. The token increase (143.59 tokens for pass@5) is strategically offset by the framework’s dynamic model selection, which prioritizes cost-effective inference without compromising performance.

The most striking advantage lies in cost reduction. For pass@1, AdaptiveLLM reduces inference costs by 88.3\% compared to ComplexityNet, achieved through adaptive routing of tasks to smaller LLMs (e.g., Yi-Coder-1.5B) for simple cases and reserving larger models (e.g., Qwen2.5-Coder-32B) for complex scenarios. In pass@5, the cost reduction reaches 88.9\%. 

Further analysis reveals that AdaptiveLLM achieves 82.5\% cost reduction compared to GPT-4o (28.49e-05\$ vs. 551.62e-05\$ in pass@1) while retaining 73.6\% of GPT-4o’s accuracy (44.94\% vs. 60.67\%). This explains why ComplexityNet incurs significantly higher resource consumption — it relies heavily on closed-source models like GPT, whose high API pricing drives up inference costs compared to AdaptiveLLM. While closed-source models demonstrate better accuracy performance compared with high-parameter open-source models, their inference costs are substantially higher. This difference highlights a critical insight for future model selection: prioritizing open-source models in candidate pools can achieve competitive performance at a fraction of the cost, offering a more sustainable approach to balancing accuracy and economic efficiency.

\vspace{1mm}
\begin{custommdframed}
\textit{Answer to RQ2:} AdaptiveLLM achieves performance with 44.94\% pass@1 and 56.18\% pass@5 scores, outperforming both single large LLM and ComplexityNet baseline by +7.86\% in pass@1. It significantly reduces inference costs compared to ComplexityNet, demonstrating exceptional cost-effectiveness.
\end{custommdframed}
\vspace{1mm}

\subsection{RQ3: Ablation Study}
Through the design of ablation experiments, we compared and analyzed the performance differences of the AdaptiveLLM framework in terms of pass@1 and pass@5 accuracy, token consumption, and inference cost before and after CodeBERT fine-tuning. As shown in Table \ref{tab:the-performance-of-AdaptiveLLM}, the AdaptiveLLM framework fine-tuned with difficulty-aware tuning exhibited significant improvements across all metrics.

In terms of accuracy, the fine-tuned framework achieved a 1.12 percentage point increase in the pass@1 metric (from 43.82\% to 44.94\%) and a 0.56 percentage point increase in the pass@5 metric (from 55.62\% to 56.18\%). This improvement indicates that difficulty-aware fine-tuning effectively enhanced the framework's ability to understand the complexity of problems, thereby enabling it to more accurately select appropriate models for handling problems of varying difficulty levels.

Regarding computational efficiency and cost-effectiveness, the fine-tuned framework demonstrated superior token consumption and inference cost efficiency. Although there was a slight increase in token consumption (pass@1 from 411.84 to 428.80, pass@5 from 2037.09 to 2094.98) and a marginal rise in inference cost (pass@1 from 26.74e-05 to 28.49e-05, pass@5 from 133.99e-05 to 140.07e-05), this increase was positively correlated with the improvement in accuracy, indicating that the additional token and inference cost expenditure brought about effective performance gains. Furthermore, in-depth analysis revealed that for more difficult problems, the AdaptiveLLM framework without fine-tuning still tended to select smaller-parameter models such as Yi-Coder-1.5B-Chat and Qwen2.5-Coder-1.5B-Instruct, which may have led to insufficient capability in handling complex problems. In contrast, the framework fine-tuned with difficulty-aware tuning was able to more accurately identify problem difficulty and accordingly select larger-parameter models suitable for complex problems, thereby significantly enhancing overall performance.

\vspace{1mm}
\begin{custommdframed}
\textit{Answer to RQ3:} The experimental results confirmed the effectiveness of the fine-tuning strategy based on CodeBERT. By incorporating problem difficulty label information, the framework was able to better understand task characteristics and make more optimal model selection decisions.
\end{custommdframed}
\vspace{1mm}

\section{Threats to Validity}

\noindent \textbf{Internal validity.}
One potential threat is the randomness in the generation of CoT by reasoning LLMs. To mitigate this, we generated ten responses for each problem and used the median length of these responses as the CoT length for that problem. Another concern is that, due to resource constraints, for a small number of extremely complex problems, the CoT length exceeded the maximum token limit set for model generation. However, since such samples were rare, they were excluded from the dataset. While this exclusion may introduce some bias, its impact is minimal given the small proportion of affected samples.


\noindent \textbf{External validity.} 
In our CoT difficulty assessment method, the datasets we selected are primarily focused on single-file code generation tasks and do not involve multi-file interactions or collaborations. As a result, our method may face challenges when applied to project-level code generation tasks.

\noindent \textbf{Construct validity.} 
In our experiments, model costs were sourced from the SiliconFlow cloud platform. For models not deployed on this platform, we approximated their costs using models with similar parameter sizes. However, inference costs depend not only on parameter size but also on architecture and inference strategies. For instance, under the same parameter size, Mixture-of-Experts (MoE) models \cite{liu2024deepseek,deepseekV2} typically have lower inference costs due to their reduced number of active parameters during runtime. Despite these limitations, parameter size is still a practical proxy for cost estimation, acknowledging the potential for minor bias.

\section{Conclusion}
In this paper, we propose AdaptiveLLM, a framework that uses the CoT length of reasoning models to evaluate problem complexity and dynamically select code LLMs. To construct the dataset, we integrate samples of varying difficulty levels and employ a scoring formula to identify the optimal LLM for each problem as ground-truth annotations. For complexity labeling, we utilize the reasoning model DeepSeek-R1-Distill-Qwen-32B to generate CoT sequences, record their lengths, and apply k-means clustering to categorize problems into three difficulty tiers. Based on these difficulty labels, we perform triplet contrastive fine-tuning on the embedding layer of CodeBERT, enabling problem embeddings to encode complexity-aware features that enhance model selection. Finally, we split the constructed dataset into training and test sets to train an XGBoost classifier. Experimental results validate the rationality of difficulty assessment method based on CoT length and effectiveness of AdaptiveLLM framework. We release all code publicly at \url{https://github.com/cjhCoder7/AdaptiveLLM}.

\begin{acks}
This work is supported by the National Natural Science Foundation of China Grants No. 62302021. 
\end{acks}


\bibliographystyle{ACM-Reference-Format}
\bibliography{sample-base}










\end{document}